# USING THE TRANSVERSE DIGITAL DAMPER AS A REAL-TIME TUNE MONITOR FOR THE BOOSTER SYNCHROTRON AT FERMILAB*

N. Eddy[#], O. Lysenko, Fermilab, Batavia, IL 60510, U.S.A.

*Abstract*

The Fermilab Booster is a fast ramping (15Hz) synchrotron which accelerates protons from 400MeV to 8GeV. During commissioning of a transverse digital damper system, it was shown that the damper could provide a measurement of the machine tune throughout the cycle by exciting just 1 of the 84 bunches with minimal impact on the machine operation. The algorithms used to make the measurement have been incorporated into the damper FPGA firmware allowing for real-time tune monitoring of all Booster cycles.

## INTRODUCTION

The Fermilab Booster synchrotron operates over an energy range of 400 MeV to 8 GeV at 15 Hz cycle. To accomplish this the RF sweeps from 37MHz to 52.8MHz. The Booster has a harmonic number of 84 and typically accelerates 81 bunches with a three bunch "notch" for the kicker. The design betatron frequencies in the horizontal and vertical planes are $v_x = 6.75$ and $v_y = 6.85$ respectively but vary substantially over a machine cycle due to lattice functions changing as the magnets ramp. A correction-magnet assembly consisting of a horizontal and vertical dipole, a quadruple and a skew quadruple is placed in each short and each long straight section. The quadruples and skew quadruples are designed to accommodate the space-charge tune shift at injection and to control the tune against inherent resonances and the coupling resonance of the horizontal and vertical oscillations over the entire cycle. Historically tune control in the Booster has been a difficult job because the tune measurement is slow and generally invasive requiring dedicated study periods [1,2].

## BOOSTER TRANSVERSE DAMPER

A digital feedback system is being commissioned to provide wideband bunch by bunch transverse damping in Fermilab Booster. An overview of the digital damper system is shown in Fig. 1. The system consists of a stripline BPM and difference hybrids which provides a signal proportional to the beam position, a digital damper board which processes the signals and provides an output kick for each bunch. The output kick is amplified and driven to stripline kickers.

The damper system is configured and controlled by the central processor unit. New programs for the VME CPU and the FPGA can be loaded in this way, and the per-state configuration of the damper is accessible via the standard controls system, ACNET [3]. One may trivially change "live" which bunches are to be damped, the phase advance of the transverse FIR filter and so on.

### Digital Damper Board

The digital damper is a custom designed VME board developed at Fermilab specifically for digital feedback applications. It was first used in the Fermilab Recycler transverse damper system [4]. The board contains four 12 bit 212Msps AD9430 ADCs input to the FPGA which drives four channels of 14 bit 212Msps AD9736 DACs. The FPGA is an Altera Stratix II EP2S60. The FPGA handles all signal processing and I/O on the board including the VME slave interface.

A schematic diagram of the data processing used for the Booster bunch by bunch damping is shown in Fig. 2. The damper board locks to the Booster RF so that the

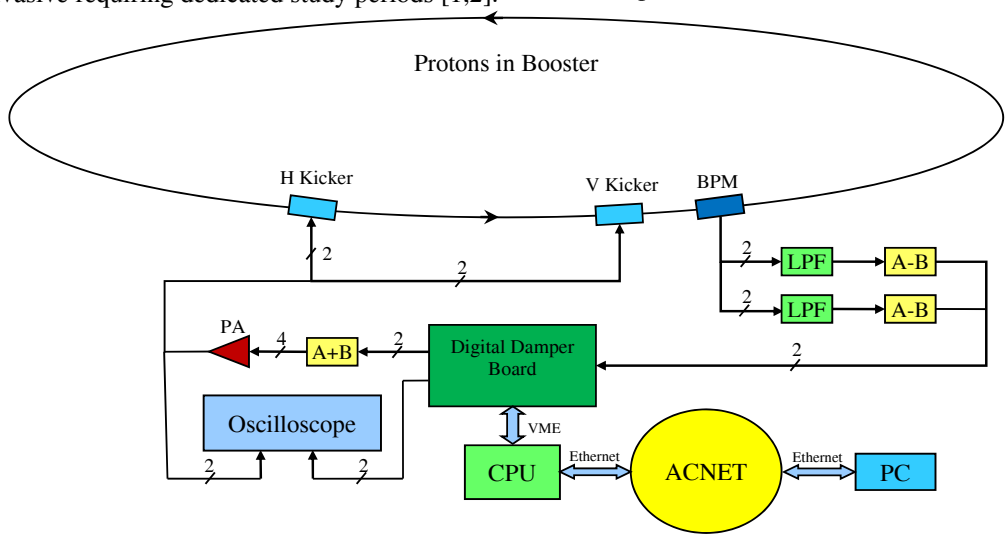

Figure 1: Overview of the Booster transverse digital damper system.

ADC, digital logic, and DACs are all synchronized with the beam RF as it sweeps. The sampled data from the ADCs input goes to the digital down convertor. The system is phased such that the *Q* component is proportional to the position. A five turn filter on each bunch position is implemented to determine the necessary bunch kick. The filter coefficients are set to remove the closed orbit position and account the pickup to kicker phase advance. Programmable gain and delay are applied to the resultant kick. Both the gain and delay change throughout the cycle to account for the energy and frequency sweep as the Booster ramps.

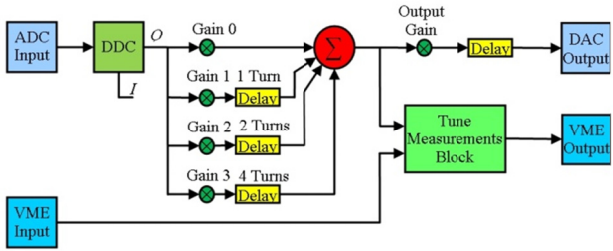

Figure 2: Block diagram of the damping algorithm.

*Tune Monitor*

The tune monitor was implemented in the firmware as an independent block which acts in concert with the damper. The horizontal and vertical measurements are independent and typically operate on different bunches. In each case, a select bunch is excited then the monitor uses a fast Fourier transform algorithm to measure the tune from position data of the excited bunch over 128 turns. The monitor makes periodic measurements throughout the Booster cycle to track the tune.

The tune monitor is able excite a bunch either by applying random noise, anti-damping, or a combination of the two. In the case of the noise, the excitation is controlled simply by the number of turns. For anti-damping, it is number of turns and gain. Up to 64 excitations or measurements can be made throughout the cycle with the each excitation independently controlled. As the damper makes many small kicks, rather than a large single turn kick, it can be better controlled and tailored to machine response at a specific time in the Booster cycle.

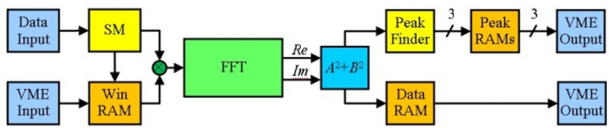

Figure 3: Block diagram of the tune measurement.

A block diagram of the algorithm to measure the tune from the bunch position is shown in Fig. 3. All design components are clocked at the Booster RF frequency. The input parameters include the selected bunch, window function, and tune limits of the peak search algorithm. The state machine then controls which bunch position to pass to the FFT as well at which times to measure it. The FFT block has a transformation length of 128 and works in the burst mode – once bunch positions for 128 turns are collected, the block performs the FFT calculation and outputs the real and imaginary parts of the spectrum from which the magnitude is calculated. In addition a peak finder algorithm examines the spectra of each measurement and stores the three highest peaks within a selected tune window. The stored tune spectra and tune peaks are available via the VME interface.

## RESULTS

The turn by turn positions for the target excited bunch and a preceding unperturbed bunch are shown for the vertical excitation of 16 turns in Fig. 4 and the horizontal excitation of 128 turns in Fig. 5. Due to horizontal chromaticity and to a lesser extent the smaller $\beta$ function at the horizontal kicker, it is much more difficult to excite the beam in horizontal plane. In fact, Fig. 5 shows clearer vertical growth. This is due to coupling and the large horizontal chromaticity.

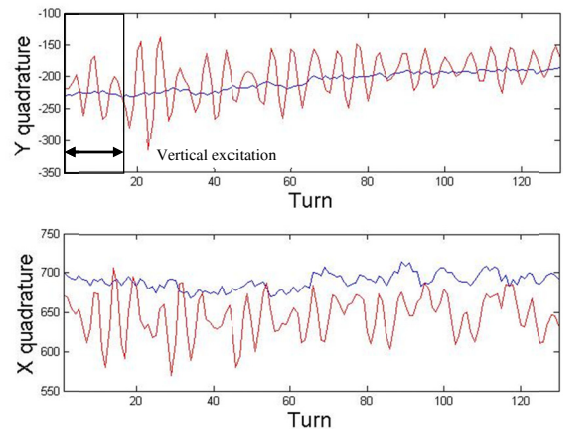

Figure 4: Turn by turn positions for the vertically excited bunch (red) and an unperturbed bunch (blue).

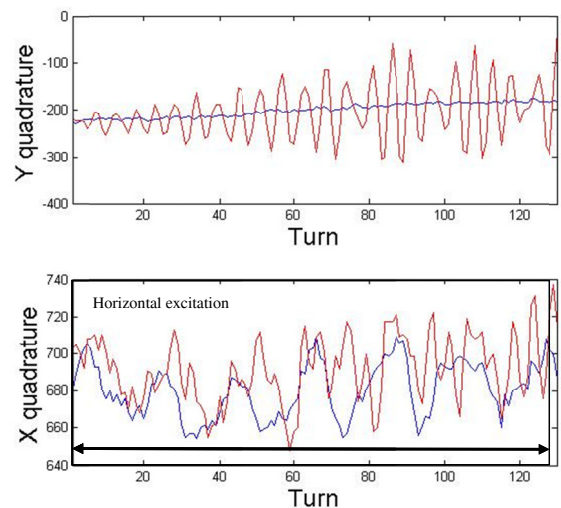

Figure 5: Turn by turn positions for the horizontally excited bunch (red) and an unperturbed bunch (blue).

The results of the FFT tune spectrum for the vertically excited bunch are shown in Fig. 6. The vertical tune is clearly present on the excited bunch as well as the horizontal tune due to coupling. There is no activity on the preceding bunch.

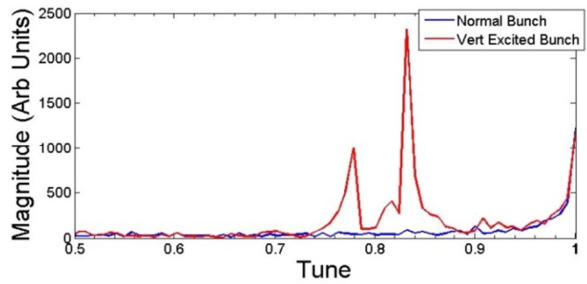

Figure 6: Tune spectra from single excitation.

To display the tune response across the full Booster cycle, a contour plot is generated from the spectrum data. An example result from the ACNET interface is shown in Fig. 7. Without normalization, the horizontal response is too small to see during much of the cycle. By normalizing each measurement spectra individually, it is possible to see the horizontal tune response even when it is very weak as shown in Fig. 8.

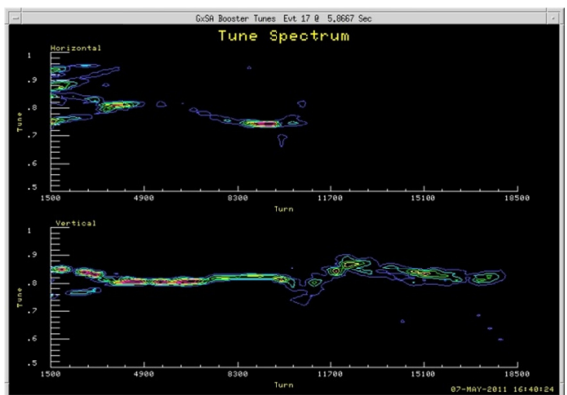

Figure 7: Tune spectrum in ACNET.

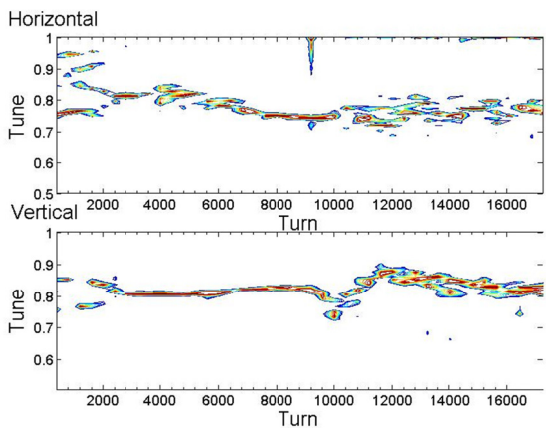

Figure 8: Normalized tune spectrum to enhance small response.

It takes about 1 second to readout and display the full tune spectrum. As noted, the algorithm also makes the 3 highest peaks in each spectrum available. This small amount of data shown in Fig. 9 can be readback and displayed at the full Booster repetition rate of 15Hz.

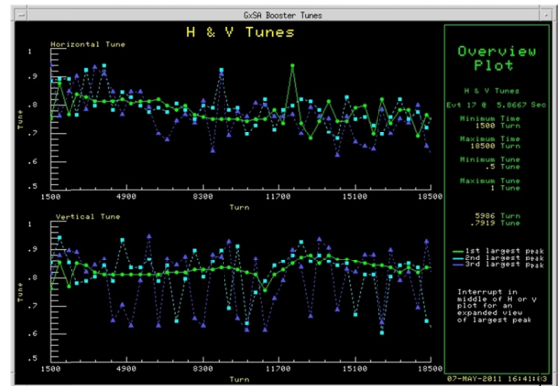

Figure 9: Horizontal and vertical tune peaks in ACNET.

## CONCLUSIONS

A new Booster tune monitor was implemented in the digital damper which has minimal impact on the Booster operation. The tune measures the tunes in two planes over the energy ramping cycle with an accuracy of 0.01 in real time.

## ACKNOWLEDGEMENTS

The authors would like to thank Salah Chaurize for invaluable help with the Booster operations.